%% file: main.tex
\documentclass[preprint, journal]{vgtc}               

\ifpdf
  \pdfoutput=1\relax                   
  \usepackage{graphicx}                
  \DeclareGraphicsExtensions{.pdf,.png,.jpg,.jpeg} 
\else
  \usepackage{graphicx}                
  \DeclareGraphicsExtensions{.eps}     
\fi%

\graphicspath{{figures/}{pictures/}{images/}{./}} 

\usepackage{microtype}                 
\PassOptionsToPackage{warn}{textcomp}  
\usepackage{textcomp}                  
\usepackage{mathptmx}                  
\usepackage{times}                     
\usepackage{cite}                      
\usepackage{tabu}                      
\usepackage{booktabs}                  
\usepackage{amssymb}
\usepackage{tikz}
\usepackage{soul}
\usepackage{xcolor}
\usepackage{comment}

\vgtcinsertpkg

\title{ICE: An Interactive Configuration Explorer for \\ High Dimensional Categorical Parameter Spaces.}

\author{Anjul Tyagi, Zhen Cao, Tyler Estro, Erez Zadok, Klaus Mueller\thanks{e-mail: \{aktyagi, zhccao, testro, ezk, mueller\}@cs.stonybrook.edu}} %
\affiliation{\scriptsize Department of Computer Science, Stony Brook University}

\teaser{
  \centering
  \includegraphics[width=\linewidth]{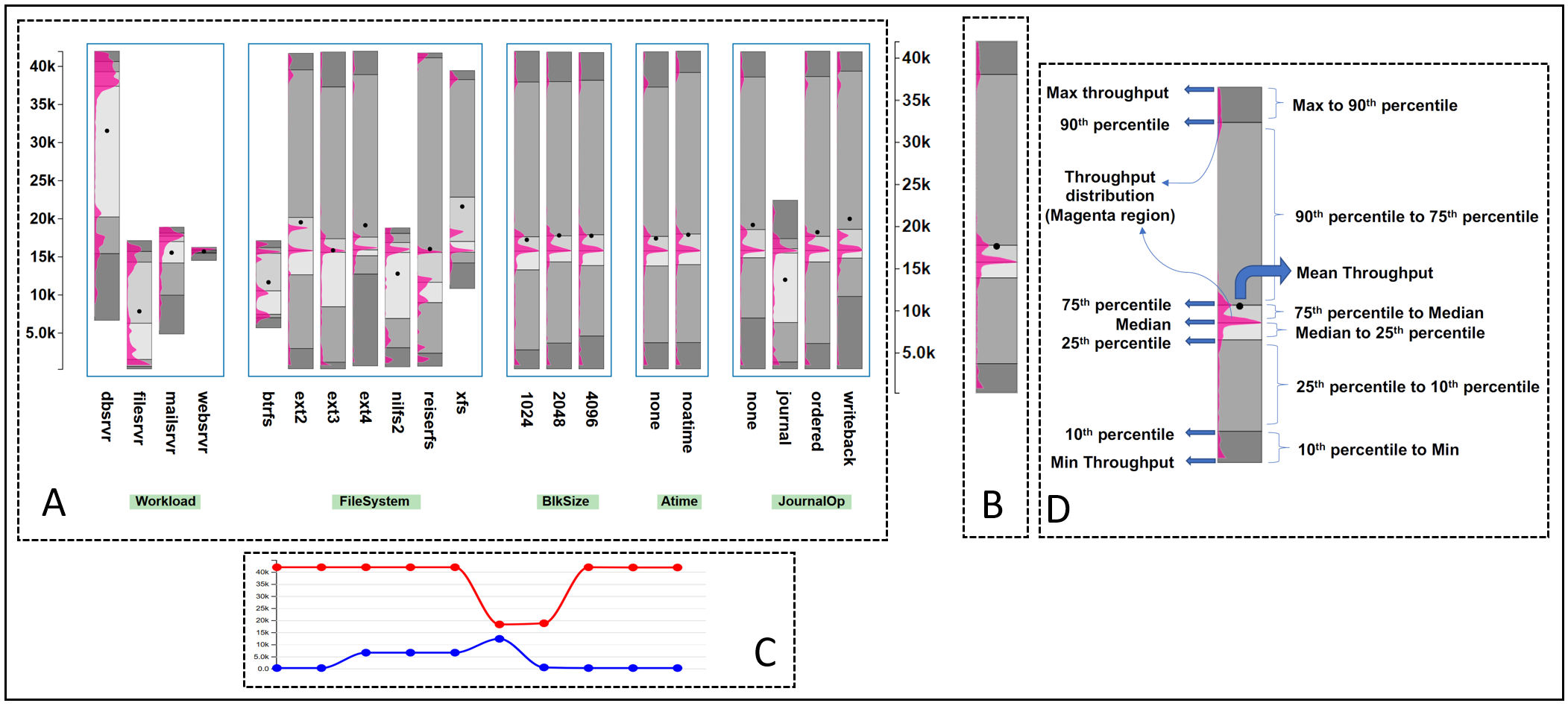}
  \caption{\textbf{A B C:} Interface of our Interactive Configuration
    Explorer (ICE) tool used to explore high dimensional parameter spaces.
    This example shows the use of the ICE in a computer systems performance
    optimization scenario.  \textbf{A} is the Parameter Explorer.  It shows
    the distribution and statistics of the numerical target variable in the
    context of the various categorical variables (or parameters), labeled by
    the green buttons at the bottom of the interface (e.g., Workload, File
    System).  Each parameter has levels e.g., Workload has 4 levels (dbsrvr,
    filesrvr, mailsrvr, and websrvr), and each level has an associated bar
    displaying the statistical information about the numerical target
    variable (here, system throughput) for this level.  Analysts can
    interactively deselect (and select) parameter levels to filter out the
    associated parameter configurations throughout.  \textbf{B} is the
    Aggregate View, which visualizes the joint distributions of all
    currently selected parameter levels.  \textbf{C} is the Provenance Terminal,
    to keep track of the changes in the target variable
    over the course of the user interactions.  \textbf{D }shows the
    information contained in each bar inside the Parameter Explorer and
    Aggregate View.}
  \label{fig:teaser}
}

\abstract{There are many applications where users seek to explore the impact
  of the settings of several categorical variables with respect to one
  dependent numerical variable.  For example, a computer systems analyst
  might want to study how the type of file system or storage device affects
  system performance.  A usual choice is the method of Parallel Sets
  designed to visualize multivariate categorical variables.  However, we
  found that the magnitude of the parameter impacts on the numerical
  variable cannot be easily observed here.  We also attempted a dimension
  reduction approach based on Multiple Correspondence Analysis but found
  that the SVD-generated 2D layout resulted in a loss of information.  We
  hence propose a novel approach, the \emph{Interactive Configuration
    Explorer} (ICE), which directly addresses the need of analysts to learn
  how the dependent numerical variable is affected by the parameter
  settings given multiple optimization objectives.  No information is lost as ICE shows the complete distribution
  and statistics of the dependent variable in context with each categorical
  variable.  Analysts can interactively filter the variables to optimize for
  certain goals such as achieving a system with maximum performance, low
  variance, etc.  Our system was developed in tight collaboration with a
  group of systems performance researchers and its final effectiveness was
  evaluated with expert interviews, a comparative user study, and two case
  studies. }

\ieeedoi{10.1109/VAST47406.2019.8986923}

\CCScatlist{
  \CCScatTwelve{Data Clustering}{Illustrative Visualization}{User Interfaces}{High Dimensional Data};
}

\newcommand{\Cross}{\mathbin{\tikz [x=1.4ex,y=1.4ex,line width=.2ex] \draw (0,0) -- (1,1) (0,1) -- (1,0);}}%

\begin{document}


\firstsection{Introduction}
\input{introduction.tex}
\input{related_work.tex}
\input{dataset.tex}
\input{requirements.tex}
\input{ICEintro.tex}
\input{variable_explorer.tex}
\input{timeline_view.tex}
\input{aggregate_view.tex}
\input{interaction.tex}
\input{design_alternatives.tex}

\input{implementation.tex}
\input{evaluation.tex}
\input{conclusion.tex}
\input{ack.tex}

\bibliographystyle{abbrv-doi}

\bibliography{template}
\end{document}

%% file: introduction.tex
\maketitle

Visual analytics of multivariate categorical data with numerical dependent
variables is crucial in many different applications, including survey
analysis~\cite{chen2015survey}, road accidents analysis~\cite{tyagiroad},
customer feedback analysis~\cite{wu2010opinionseer}, and computer systems
performance research~\cite{tarasov2011benchmarking, chen2017vnfs, cao2019graphs}.  To study
and compare categorical variables, we often need to understand their
behavior with respect to one or more numerical variables, as numerical
variables have well-defined statistical meaning and hierarchy.  For example,
in a road accidents study, the categories (\textit{Monday, Tuesday, etc.})
of the variable (\textit{day of accident}) can be correlated by studying the
number of accidents on each day.  Similarly for computer systems performance
analysis, the configurations of the categorical variable (\textit{hard disk
  types}) can be compared by studying their effects on the system's
throughput.  Sedlmair \textit{et al.}~\cite{sedlmair2014visual} defined six
analysis tasks that often recur in similar parameter spaces: optimization,
partitioning, outliers, fitting, sensitivity and uncertainty.  Our objective
is to support optimization, partitioning and sensitivity analysis of the parameter space with an expressive visual interface. ICE can be used to analyze the spread of the dependent numerical variable with respect to every parameter. Also, the parameter space can be partitioned with interactive filtering based on user goals.

Most existing parameter-visualization methods decompose a high-dimensional
space into a matrix of small multiples, each showing the relation among two
parameters.  Some researchers use bivariate scatter-plot
projections of the full space while others use HyperSlices, a set of orthogonal 2D slices,
each holding the target configuration as a center focal point~\cite{berger2011uncertainty,
  piringer2010hypermoval}.  The
shortcoming of such methods is that they only show two parameters per plot,
turning the quest for insight about multivariate relationships into a visual
search across the plots, requiring mental fusion of disjoint
relationships. Also, only a few techniques exist for analyzing the parameter
spaces of categorical variables, such as Parallel
Sets~\cite{kosara2006parallel} and
SVD-based displays generated by Multiple Correspondence
Analysis~\cite{greenacre1984correspondence}
These visualization techniques can be classified mainly into two types: (1)
dimension-reduction techniques for categorical data and (2) data splitting
based on categorical features.  Both techniques suffer from certain shortcomings,

One of these shortcomings is information loss. For techniques based on MCA and similar dimension reduction procedures, the generated layout suffers from information loss. For complex datasets, parameter relationships might not be preserved in lower dimensions, which can result in a misinterpretation of the parameter space.

%

Another shortcoming is that the existing techniques are not overly well suited for visually optimizing multiple objectives at the same time.
Consider a systems engineer who wants to filter configurations based on high
throughput and small throughput variance simultaneously. These two user goals in this case are the objectives for searching through the parameter space which have to be optimized simultaneously. Visualizing the
parameter space in context of the dependent numerical variable for multiple
objectives is not possible with dimension-reduction techniques.  Parallel
Sets, on the other hand, allow for multi-objective filtering but the
polylines or sectors can become too cluttered as the number of variables and
levels in the dataset increases.
%

We collaborated with a group of computer systems
researchers who faced exactly these challenges. We began with assessing the requirements of an effective visualization tool
that would effectively enable them to study a set of categorical variables in context of a numerical dependent
variable in light of multiple optimization objectives. Based on an analysis of these requirements we then iteratively derived a novel approach for this purpose, called the \textit{Interactive
Configuration Explorer (ICE)} that is subject of this paper.
%

ICE is a tool especially designed for tuning a large number of categorical
parameters, for objectives based on a dependent numerical variable, like in
computer system performance optimization~\cite{cao2018towards} where the
objective is based on the throughput behavior of the system. One of the important reasons for developing ICE is to assist the analyst in visualizing the search space at every stage in the optimization process. Hence, the parameters are visualized based on the range and distribution of the dependent numerical variable they span. This representation is free of any information loss because the categorical variables are not transformed into numerical variables but are studied as individual identities, hence preserving the properties for both ordered and unordered categorical variables. We evaluate ICE
for performance, effectiveness and generality with the
help of two case and two user studies. The main contributions of our work
are:

\begin{itemize}
\setlength{\itemsep}{-2pt}
    \item Visualize a greater number of categorical variables with a view facilitating comparison between all parameter levels.
    \item Assist in multi-objective optimization based filtering on large parameter spaces.
    \item Compare multiple configurations (set of parameters) based on their impact on the dependent numerical variable.
\end{itemize}

Our paper is organized as follows.  Section 2 presents related work.
Section 3 present the dataset and domain setting we used to gain a practical
backdrop for this otherwise rather general design.  Section 4 presents a
requirement analysis characterizing these types of applications.  Section 5
describes our methodology, the ICE, along with two case studies rooted
within the systems domain.  Section 6 presents some helpful implementation
hints.  Section 7 outlines a thorough evaluation we performed with a set of
more general case studies to show the generality of our tool.  Section 8
concludes.


%% file: related_work.tex
\section{Related Work}
\label{s:related}

%
%
In this section, we will discuss the existing techniques available for studying mixed multivariate datasets including both
categorical and numerical attributes applied in related domains~\cite{waser2010world,heinzl2017star}. The main
objectives of visual analytics in these domains includes the study of correlations between categorical variables and clustering in the parameter space with projection methods (fused displays and dimension reduction techniques) or parallel sets.

\subsection{Techniques to study correlation}

There are multiple specialized techniques available to study correlation
between features in high-dimensional data.  Since the data in consideration is categorical
with one dependent numerical variable, most techniques like Pearson
correlation will give ambiguous results.  Hence, specialized correlation
methods like Cramer's V (based on Chi-squared statistic) are
used~\cite{batagelj2004pajek, epskamp2012qgraph}.  There also exist statistical tests for correlating categorical variables by comparing their
behavior on numerical variables, like T-test, chi-square test, One-Way ANOVA
and the Kruskal Wallis test. Techniques also exist to study correlation of multivariate temporal data \cite{cappers2017exploring, unger2017understanding}. However, for datasets with very high dimensionality, it can be hard to study correlations in the overall distribution of the dataset. Hence, methods to study correlation on large datasets over parts of the distribution have been devised~\cite{shao2017interactive}. The results from these techniques can then be used as input
to fused displays where these correlations are visualized in the form of
scatter-plots and networks~\cite{zhang2012network}.
%
%

\subsection{Clustering techniques}

Since most categorical data consist of unordered nominal
values~\cite{zhao2017clustering}, most clustering algorithms are not
directly applicable to study categorical parameter spaces.  Advanced techniques like
k-mode~\cite{huang1998extensions},
SQUEEZER~\cite{he2002squeezer} and COOLCAT~\cite{barbara2002coolcat} have
been developed to work especially on categorical data.  Some of the latest
research has focused more on advanced clustering techniques in a supervised
learning environment~\cite{wang2018perception} based on human perception. All of these techniques differ based on the similarity criterion used for clustering as different similarity criterion are designed to capture specific relationships in the data. However, in multi-objective filtering scenarios, clustering as a concept is limited in its scope as each algorithm captures only a particular relationship in the dataset based on the similarity criterion.
%

\begin{figure}[tb]
 \centering
 \includegraphics[width=\columnwidth]{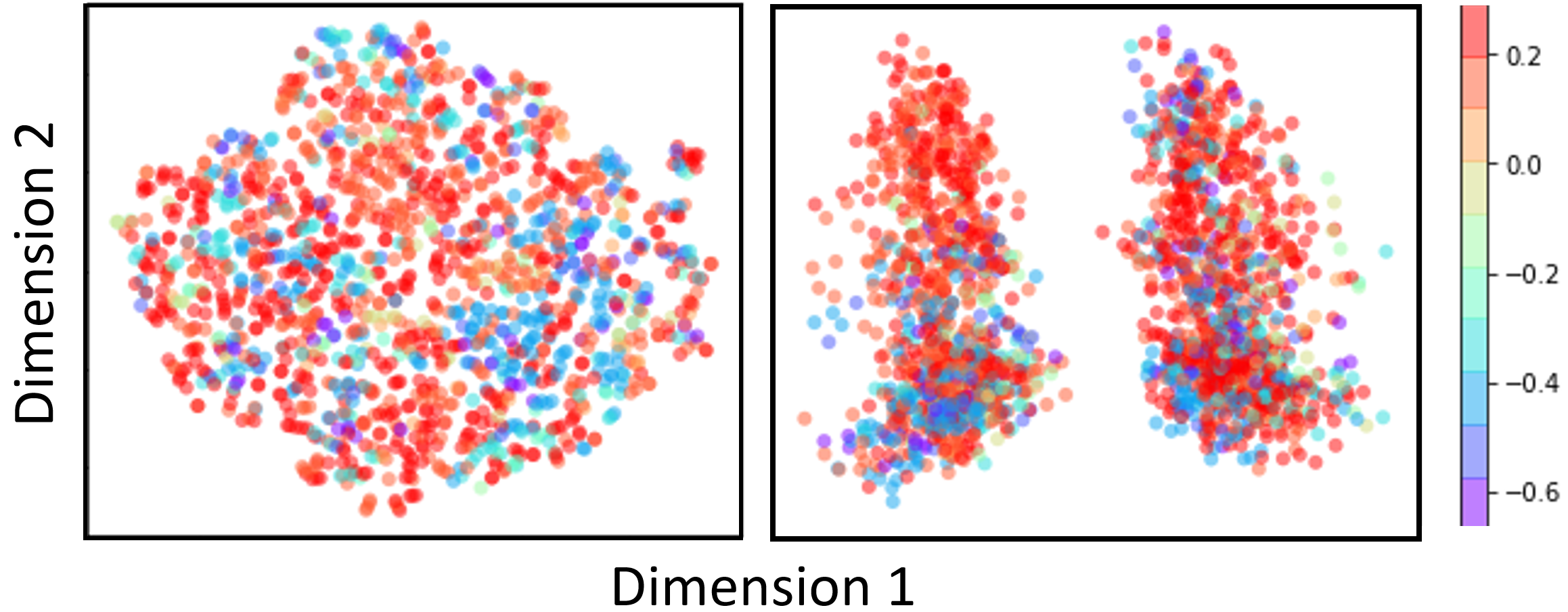}
 \caption{Visualizing our systems performance dataset with t-SNE(left) and spectral clustering(right).  Each datapoint is projected into two dimensions
   and the color of a point represents the throughput value on the
   normalized scale from -1 to 1.  The objective is to visualize the
   clusters of throughput values.  But no clusters with respect to the
   throughput could be seen as the values are spread uniformly across the
   plot.}
 \label{fig:spec_clustering}
\end{figure}

\begin{figure*}[th]
    \centering
        \includegraphics[width=\textwidth]{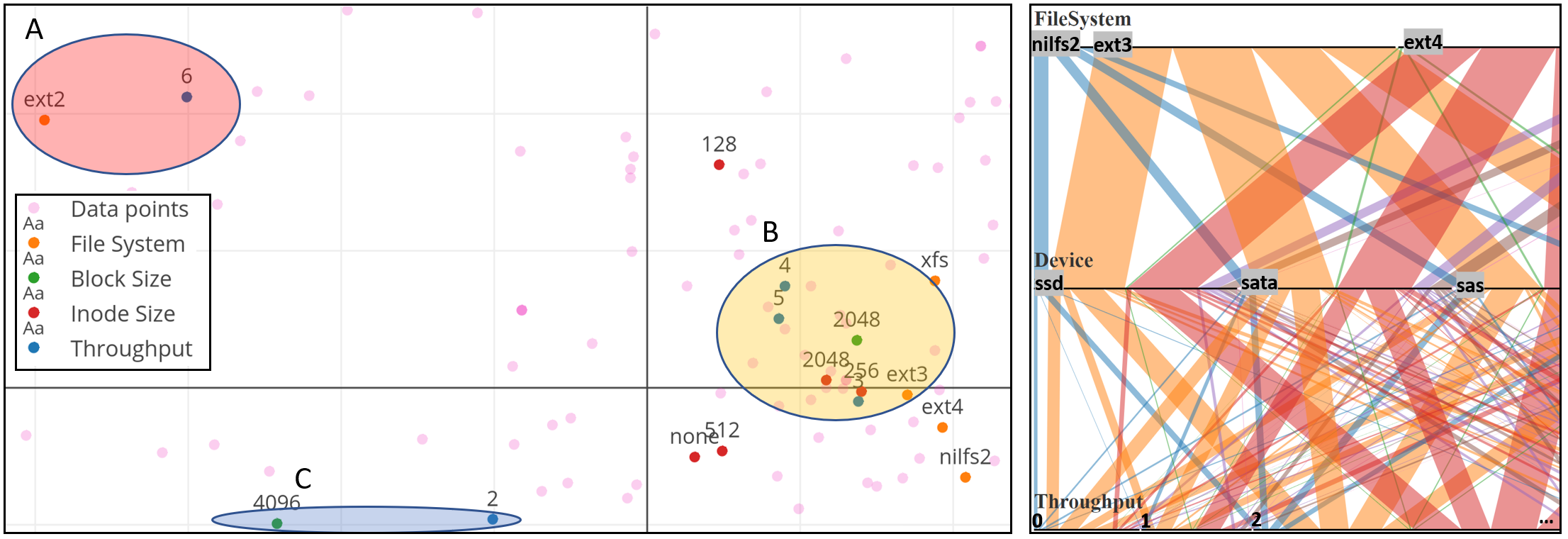}
    \caption{\textbf{Left:} MCA plot of our system performance dataset. System throughput has been discretized into six categories labelled as numbers from 1 to 6 on the plot with blue nodes.
      \textbf{A} shows the high throughput region and ext2 (a File System parameter level)
      is the closest level, i.e., most correlated with the highest
      throughput region (the blue node labeled '6').  \textbf{B} shows the
      moderate-high throughput region (the blue node labeled '4') with block
      size 2,048 being the most correlated level.  Similarly, \textbf{C}
      shows that block size = 4,096 is most correlated to the low
      throughput region (the blue node labeled '2').  \textbf{Right:}
      Parallel sets displaying the system performance data with five
      categorical variables.  Throughput is the dependent numerical variable
      which we converted into a categorical variable via equi-width binning.
      The polylines show what percentage of data belongs to what parameter
      settings.  It is difficult to gain any insight into the dataset as the
      plot gets too cluttered with more variables.}
    \label{fig:mca_ps}
\end{figure*}

\subsection{High dimensional Data Visualization techniques}
\label{bg:projection}

Projecting high dimensional data into lower dimensions is another technique
to visualize relationships between attributes and the data points.  Scatter-plot
matrices~\cite{hartigan1975printer} is a way to visualize pairwise
relationships between the variables in which multiple plots are generated
where each plot compares two attributes from the dataset.  Different
variations of this technique include bivariate scatter-plot projections of
the full space and HyperSlices based
approach~\cite{berger2011uncertainty, piringer2010hypermoval}. However, all
of these technique do not scale with the number of attributes as the number
of plots increases exponentially. This makes it difficult to mentally fuse
the disjoint relationships obtained from individual plots. Similarly, 3D volume datasets can be represented with Multicharts~\cite{demir2014multi} and dynamic volume lines~\cite{weissenbock2018dynamic} but these techniques are also limited in their application domain.
%

Parallel Sets~\cite{kosara2006parallel} is another popular method for visual analytics of
multidimensional categorical data. It maps data into ribbons
which subdivide according to the percentage of the population they
represent. Each categorical variable is mapped to an axis which is divided
into sections according to the percentage of data contained in each category
(see Figure~\ref{fig:mca_ps} (right)).
However, as the number of parameters in the dataset increases, the plot can
become too cluttered to project any useful information.  An example parallel
sets plot of our systems performance data is shown in
Figure~\ref{fig:mca_ps} (right), showing the excessive overlap of ribbons
with only five variables.  The complete parallel sets plot is given in the
supplementary material.

Another class of dimension reduction techniques include
MDS~\cite{kruskal1977relationship, kumar2019task}, PCA, Kernel PCA, locally linear
embedding (LLE)~\cite{roweis2000nonlinear}, Fisher's discriminant
analysis~\cite{mika1999fisher}, spectral clustering~\cite{ng2002spectral} and t-distributed stochastic
neighbor embedding (t-SNE)~\cite{maaten2008visualizing}.  Although these
techniques have been designed to work with numerical data, categorical data
can be converted to numeric form and can be visualized using these
techniques.  To convert categorical data into numerical format, we can use
one-hot encoding or the re-mapping technique described by Zhang \textit{et
  al.}~\cite{zhang2015visual}.
%
These methods are good for visualizing relationships between the datapoints but their effectiveness decrease as the dimensionality of the dataset increase. An example case is shown in Figure~\ref{fig:spec_clustering} where no
clear clusters based on the dependent numerical variable (throughput) could be seen with spectral clustering and t-SNE on the systems performance dataset.

To better cater the need of projecting a larger number of dimensions to lower dimensions, another class of multi-variate projection techniques exist which arranges variables in radial layouts e.g. Star Coordinates~\cite{kandogan2000star, kandogan2001visualizing, lehmann2013orthographic} or RadViz~\cite{daniels2012properties, grinstein2001high, hoffman1997dna}. Both of these techniques are similar as they generate a radial layout with variables as anchor points on the circumference of a circle and the data points are systematically places inside the circle based on their value for each variable. Star coordinates project a linear transformation of data while RadViz projects a non-linear transformation~\cite{rubio2015comparative}. These projection techniques work well to project and visualize clusters in high dimensional numerical data~\cite{novakova2009radviz}. Also, Star coordinates and RadViz can be combined to create a smooth visual transition over multiple dimensions of the data to visualize multiple dimensions of the dataset interactively~\cite{lehmann2015optimal, lehmann2016optimal}. While these techniques work well for numerical data, they cannot be applied directly to categorical parameter spaces. A variation, concentric RadViz~\cite{ono2015concentric} can be used  to study different categorical variables as concentric RadViz circles but the main objective is to study data distribution for given parameter combinations. However, the correlation between different categories cannot be visualized with this technique.

Another technique, Multiple Correspondence Analysis
(MCA)~\cite{greenacre1984correspondence} is specifically designed for
projecting categorical data. Numerical data can also be visualized with MCA by discretizing it into categories.  It can be used to generate fused displays in which the
levels of categorical variables are plotted within the same space than the data
points.  Similar to PCA, one can select a bivariate basis which maximizes
the spatial expanse of the plot.  In these displays the distance between two
points represents a notion of association.  As shown in
Figure~\ref{fig:mca_ps} (left), MCA is effective in visualizing associations
among the levels of the categories.  However, there is a certain loss of
information due to the omission of the higher order basis vectors.  It also
tends to get cluttered when the number of data points (the parameterized
configurations) or even the number of categories and levels grow large.

%% file: dataset.tex
\section{Dataset}
\label{s:dataset}
While our method readily applies to any categorical dataset with a numerical
(or categorical) target variable, our specific use case was to support a team of systems researchers in their aim to learn about the impact of configuration
choices on throughput and its variability in a benchmark computer
system. The dataset we used had been collected over a period of three years
in the research team's lab at our university.


A set of several experiments were run to
measure the system performance for a large number of configurations.  Currently, the dataset consists of 10 dimensions with
100k configurations and about 500k data points (i.e., system configurations
that were each executed on average five times to ensure stable results).
The attributes in the dataset include \textit{Workload Type, File System,
  Block Size, Inode Size, Block Group, Atime Option, Journal Option, Special
  Option, I/O Scheduler,} and \textit{Device type.}  All of these variables
are categorical where a configuration is a set of categories from at least
one of these variables.  Some of these variables are \emph{ordinal} (e.g.,
Block Size can be 1KB, 2KB, or 4KB only) while others are \emph{nominal}
(e.g., JournalOp can be writeback, ordered, journal, or none). The dependent numerical variable is the \textit{Throughput} of each parameter configuration.

Direct optimization techniques have been applied to search for optimal configuration in such large parameter spaces. Some of the applied techniques include Control Theory~\cite{li2012power,li2011model,zhu2009does}, Genetic Algorithms~\cite{holland1992adaptation,goldberg1988genetic}, Simulated Annealing~\cite{kirkpatrick1983optimization,cohn1999simulated} and Bayesian Optimization~\cite{shahriari2015taking}. However these techniques prove to be too slow and sometimes result in sub-optimal solutions as our experiments confirm~\cite{zadok2015parametric, cao2018towards}. Hence, there is a need to visualize the search space and the efficacy of the search techniques. Our ICE tool helps in visualizing and filtering these large parameter spaces to learn about optimal settings and trade-offs for the underlying system's performance.


%% file: requirements.tex
\section{Requirement Analysis}
\label{s:req}

To systematically evolve  our ICE tool with the needs of the systems researchers in
mind, we applied Munzner's nested model for visualization
design~\cite{munzner2009nested, meyer2012four}.  Building the ICE tool
following the nested model greatly helped in the step-by-step development with
proper evaluation at each stage of the implementation.  The first of the
four stages of developing the eventual visual tool was to gather,
from the domain experts, a list of requirements expected to be met by our tool.
Our many discussions culminated in the  following list of seven
requirements:

\textbf{R1: Statistics visualization.} System researchers are typically
interested in assessing the impact of a parameter on throughput via
statistical measures.  Hence, the framework should display the \textit{Mean,
  Median,} some
  \textit{Percentiles, Min, Max,}
  \textit{Range} and
  \textit{Distribution}
  of the resulting throughput
for each variable independently. Visualizing a complete distribution curve is important to prevent any incorrect statistical information. For
example, the mean of a bimodal distribution and a normal distribution might
be the same, but they are different distributions requiring different
systems approaches to optimize.  A full distribution curve of
the data can complement the statistical information, thus preventing any
deceptive conclusions about a parameter.


\textbf{R2: Comparative visualization.} Comparing the impact and trade-offs of different
parameters on system throughput is crucial for choosing the best
configuration in such a large parameter space.
The ability to compare different parameter settings helps analysts to
determine the right set of parameters by repeated selection and filtering to
arrive at the desired system performance.

\textbf{R3: Filtering.} When dealing with large parameter spaces, choosing a
system configuration with the best performance is non-trivial.  Filtering by
choosing the best parameters iteratively can reveal complex hierarchical
dependencies between the parameters and system throughput.  For example,
assume analyst Mike seeking to optimize a system running a database server
workload.  He can first choose the best File System type, followed by the
best Block Size and so on until there is no more improvement in the system
performance.

\textbf{R4: Support informed predictions.} As discussed in R4, filtering is
important for reducing the large parameter space to a smaller space of
interest. Yet, guidelines are needed that can help an analyst choose
the right parameters to reach a desired goal.  Assume analyst Jane who has a system
running a Database server workload and a File System of type ext2.  Now she
wishes to choose the system configuration which gives a minimum variation in
the performance: i.e., the narrowest range of throughput thus yielding a
``stable'' throughput behavior.  To achieve these goals, the visualization scheme should
provide the necessary cues.

\textbf{R5: Provenance visualization.} Iterative filtering is useful but it needs
to be attached to a visual provenance scheme where the analyst can keep track of
the progress at each stage in the filtering progress. Likewise, the analyst
should be able to move back to any past state in the pipeline to undo any
actions if required.


\textbf{R6: Aggregate view.} Requirements R1-R4 focus on analyzing the
impact of throughput with each parameter in the dataset where the goal is to
assist in informed predictions. At the same time, the interface should also give a summarizing view of the span of throughput performance that is reachable with the evolving system configuration.
%

During our meetings with the systems research team, we soon realized that
they presently had very few visual tools at hand to analyze their large
parameter spaces with these seven requirements in mind. They were open to
the use of visual tools, but they strived for easy-to-understand traditional
visualization tools, as opposed to highly specialized designs with a
possibly steep learning curve. Their motivation was to develop a tool that
would gain wide acceptance within the systems-research community and use
well recognized standards and metrics, made visual and interactive via our
tool.

We also concluded that dashboards with standard visualizations, such as bar,
line, and pie charts were insufficient to fully capture the requirements we
collected, at least not in an easy and straightforward manner.  Other
visualization paradigms such as parallel sets and MCA plots were similarly
ruled out (see our study in Section~\ref{bg:projection} above).

We thus needed to find a balance between an advanced visualization design
and one that would convey the identified established performance metrics in
an intuitive way.  We believe that the emerged design and the lessons
learned throughout the process are sufficiently general and apply to domains
much wider than computer systems analysis.

%


%% file: ICEintro.tex
\section{Interactive Configuration Explorer (ICE)}

The ICE interface is divided into three components (see
Figure~\ref{fig:teaser}).  The first section is the Parameter Explorer
(\textit{A}).  Its design satisfies majority of the requirements (R1 to
R4) as it visualizes and allows users to tune the target variable's distribution
for each parameter in the dataset.  It allows the analyst to turn off
parameters that are deemed irrelevant as well as filter out configurations
with unwanted or non-competitive parameter level settings, both by toggling
on/off the parameter and parameter level (category) bars, respectively, enabling the
user to conduct the iterative optimization of the target parameter, system
throughput in this case. It also supports zooming and panning for better comparison of the bars.  To the right of the Parameter Explorer is the Aggregate View
(\textit{B}).  The Aggregate View displays the throughput distribution for
the configurations selected in the Parameter Explorer, thus satisfying
requirement R6.  The third component of the ICE is the Provenance Terminal
(\textit{C}).  It satisfies requirement R5 and allows the user to easily
track, roll back, and edit the parameter filtering progress.

\subsection{The Range-Distribution (R-D) Bars}

\begin{figure}[tb]
 \centering
 \includegraphics[width=\columnwidth]{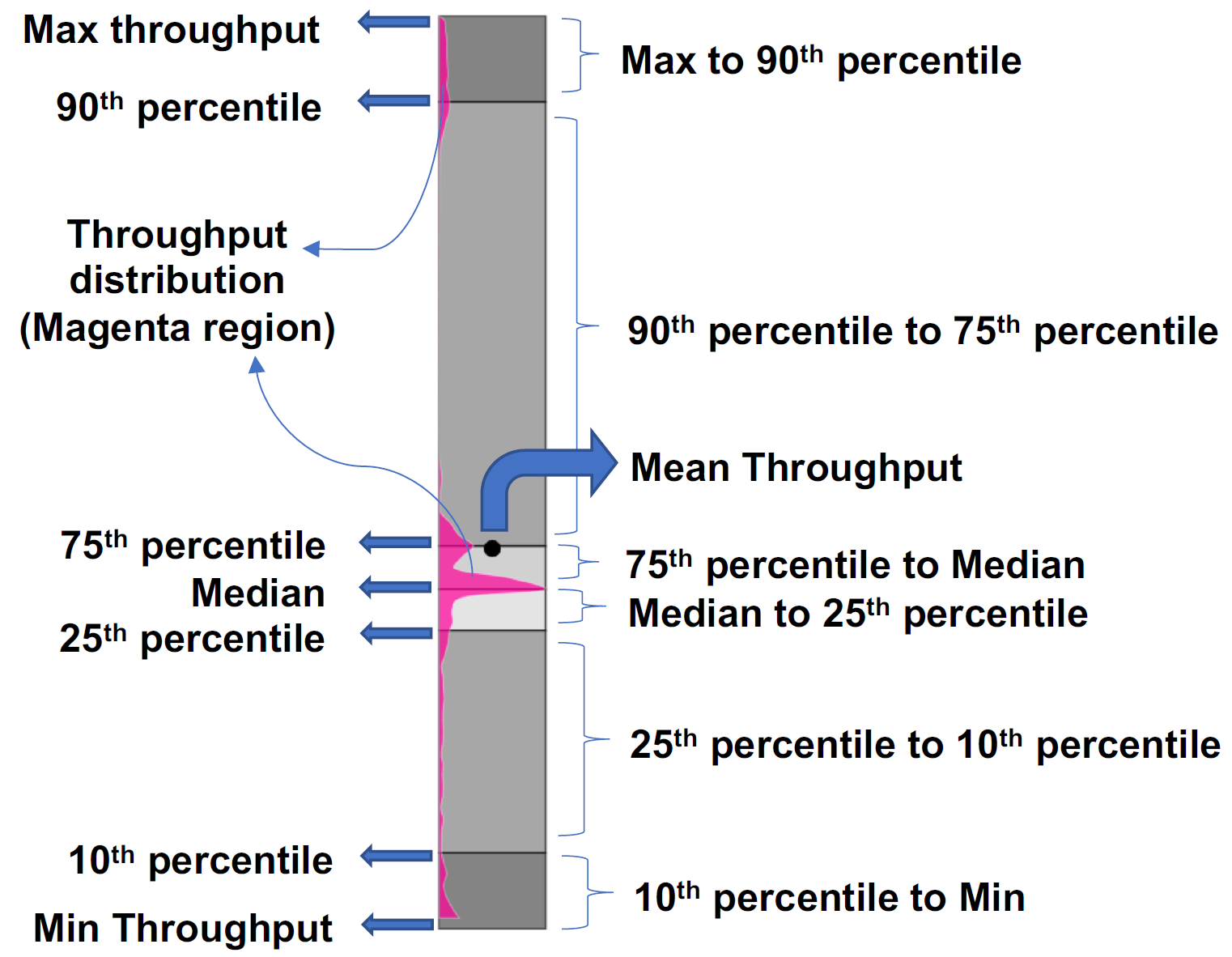}
 \caption{Annotations of Range-Distribution Bar used in the Parameter Explorer and the Aggregate
   View.  The example shows system throughput as the dependent numerical variable.}
 \label{fig:thp_annotated}
\end{figure}

Sections \textit{A, B} of the ICE interface consist of a set of
\textit{Range-Distribution (R-D) bars}.  Each bar contains the probability distribution function with additional statistical information about
the dependent numerical variable.  The R-D bars are arranged and delimited similarly to a
vertical Gantt or timeline chart, with one bar dedicated to one parameter
level, and are grouped by the variables.  The lower/upper limit of each bar is
determined by the lowest/highest value of the dependent numerical variable that can be achieved for all
configurations with the parameter level the bar represents.

A completely annotated bar displaying the
information that each part of the bar contains is shown in
Figure~\ref{fig:thp_annotated}.  Each bar is a sequence of combinations of
grays which represent the range of percentiles.  The color codes are chosen
with the help of \textit{ColorBrewer}~\cite{harrower2003colorbrewer} to show
a continuous diverging effect of percentiles on the bar.  The magenta region
shows the distribution of the target variable over the range.  Statistical
information is shown with lines separating the percentile ranges and a black
dot displaying the mean value.  See Section~\ref{s:design-alt} for more
detail on how we arrived at these specific design choices.


%% file: variable_explorer.tex
\subsection{Parameter Explorer}
\label{param-exp}

\begin{figure}[tb]
 \centering
 \includegraphics[width=\columnwidth]{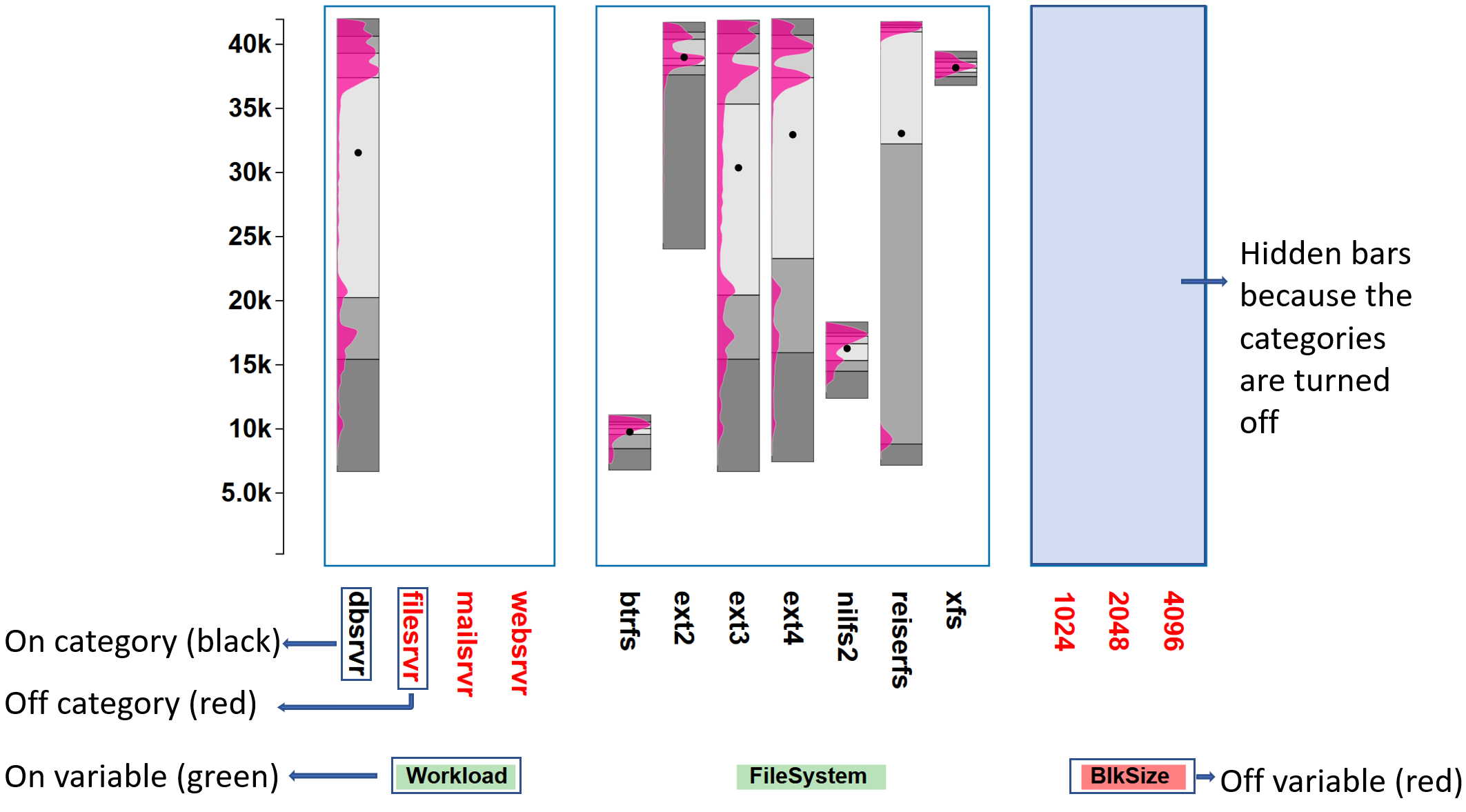}
 \caption{The Parameter Explorer in action for the system performance dataset.
   Analysts can select parameters from the Parameter Explorer and visualize
   throughput distributions and statistics in real time.}
 \label{fig:variable_explorer}
\end{figure}

The Parameter Explorer is designed for the goal of visualizing a numerical
variable with respect to individual parameters in the dataset: i.e., the
requirements R1 to R4.  As mentioned, multiple bars are stacked, grouped by parameters and their levels. This grouping allows for easy comparison of the
impact of numerical variable on the parameters.  As shown in
Figure~\ref{fig:variable_explorer}, the level names are listed underneath
each bar and the parameters are shown as buttons below the group of levels.  The bars for each variable are grouped within a blue box.  The statistics (mean and percentiles) are shown as alternating
shades of gray for each parameter level, hence partially satisfying R1.  The distribution of dependent variable is shown as a magenta distribution curve.
The grouping of bars with each bar containing the information about the
impact on the dependent variable clearly reveals the correlation between
the parameters levels, if there is any. For example, in Figure \ref{fig:teaser}, the \textit{Workload} types \textit{dbsrvr} and \textit{websrvr} can easily be compared based on the throughput values they span. A system running a \textit{wbsrvr} workload has much less variation in the throughput as compared to the system running a \textit{dbsrvr} workload. Similarly, all parameters can be correlated based on user objectives for a system optimization. This satisfies requirements R2 and R1.

Analysts can use the Parameter
Explorer to filter within a large set of possible configuration spaces.  As
shown in Figure~\ref{fig:variable_explorer}, the user has the ability to
select one or more levels for each parameter; for example, the level
\textit{dbsrvr} is selected (level name shown in black) and the remaining
levels in \textit{Workload} are not (level names shown in red).
Also, the user can completely select or remove a parameter from the dataset;
for example, \textit{Block Size} (button shown in red) is toggled off by the
analyst, so it is not considered in generating the aggregate view.  This
satisfies the filtering requirement R3.

We specifically designed the Parameter Explorer to accommodate
many parameters in a small space.  One bar is generated for one parameter level, and depending on the screen size, analysts can accommodate
several parameters in a single screen for quick comparison and filtering of
the parameter space.  Compared to parallel sets (Figure~\ref{fig:mca_ps}, right), where at the finest level one line is drawn for each data
point, or groups of identical data points (see bottom portion of the plot), the space efficiency of ICE in displaying parameter levels is highly
optimized. The simple stacked bars concept of ICE prevents the data
cluttering that plagues the parallel sets since it captures the configuration statistics succinctly in each bar.  Figure~\ref{fig:variable_explorer} shows a
portion of the Parameter Explorer for the system performance dataset.  The
complete view of the Parameter Explorer is available in the supplement material.

The analyst can click on the level label to toggle it.
Parameter Explorer and the Aggregate View are updated based on the filtered
parameter space data.  In this way, analysts can iteratively move closer to
the configurations with the desired value of the target variable,
throughput.

%% file: timeline_view.tex
\subsection{Provenance Terminal}

\begin{figure}[tb]
 \centering
 \includegraphics[width=\columnwidth]{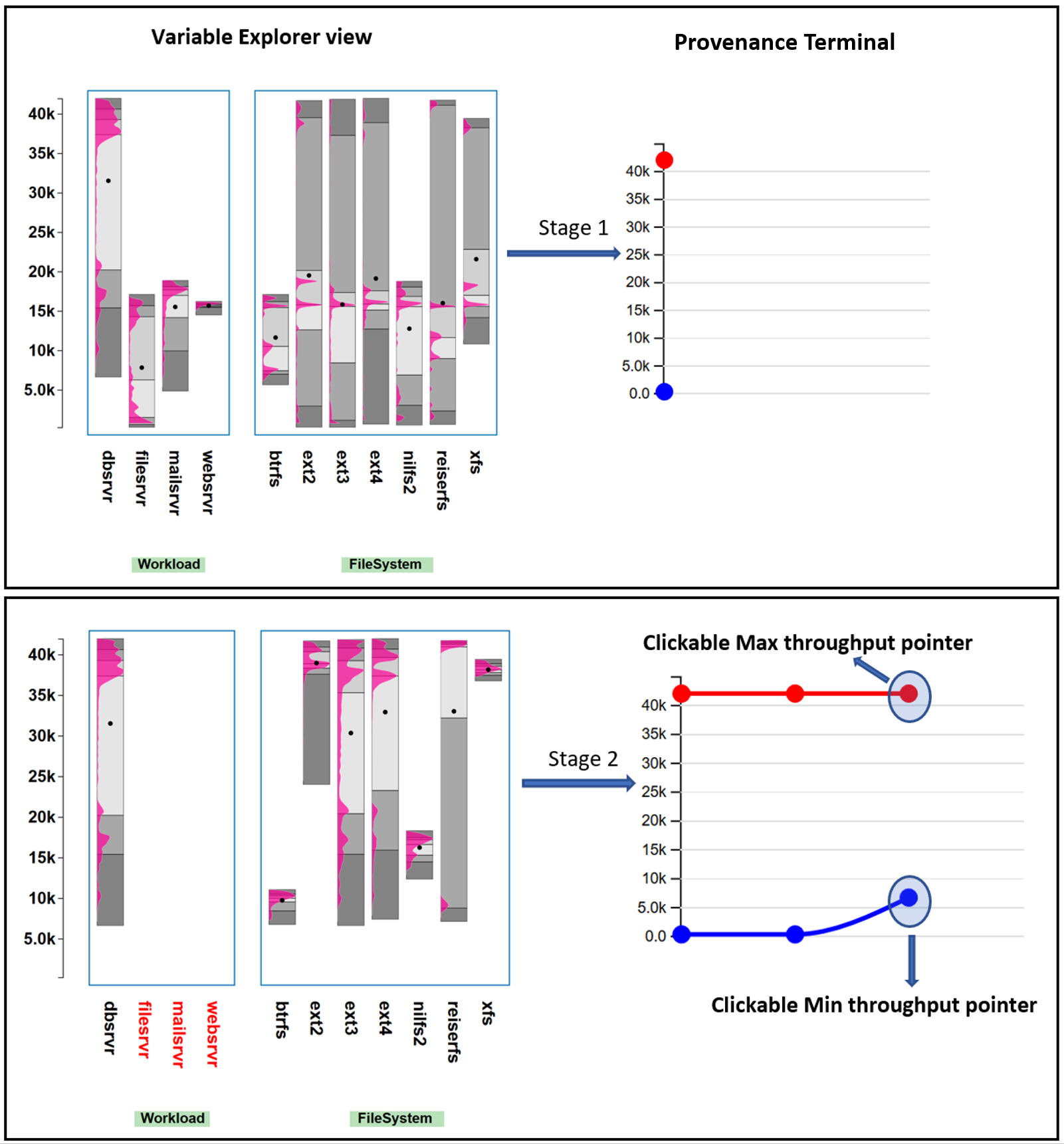}
 \caption{Provenance Terminal on the system performance dataset, showing how the
   aggregate throughput range changes with each parameter selection.  The red
   (blue) line denotes the maximum (minimum) throughput achievable with the
   current parameter settings.  The initial stage (stage 1) shows the range
   of throughput for the current overall dataset.  Stage 2 shows the updated
   provenance terminal where the analyst had selected only \textit{database server
     (dbsrvr)} as the workload type.  Each of these filtering steps can be
   rolled back by clicking on any of the pointers.}
 \label{fig:timeline_view}
\end{figure}

The Provenance Terminal (see Figure~\ref{fig:timeline_view}) is used to keep track
of the progress of the iterative filtering activities.
In this process, the analyst might want to toggle between multiple parameter
configurations to compare the resulting dependent variable
distributions.  The Provenance Terminal can be used to see and compare the
dependent variable ranges for the various iterated parameter configuration.
It also allows the analyst to roll back to a previous parameter
configuration if the evolution gets stuck without hope to further improve
it.  This satisfies requirement R6.  The maximum value of the dependent
variable at each stage of the selection is shown with a red circular pointer
on a red line, while the minimum value is shown with a blue circular pointer
on a blue line.  This view is updated with each user interaction.

\begin{figure}[tb]
 \centering
 \includegraphics[width=\columnwidth]{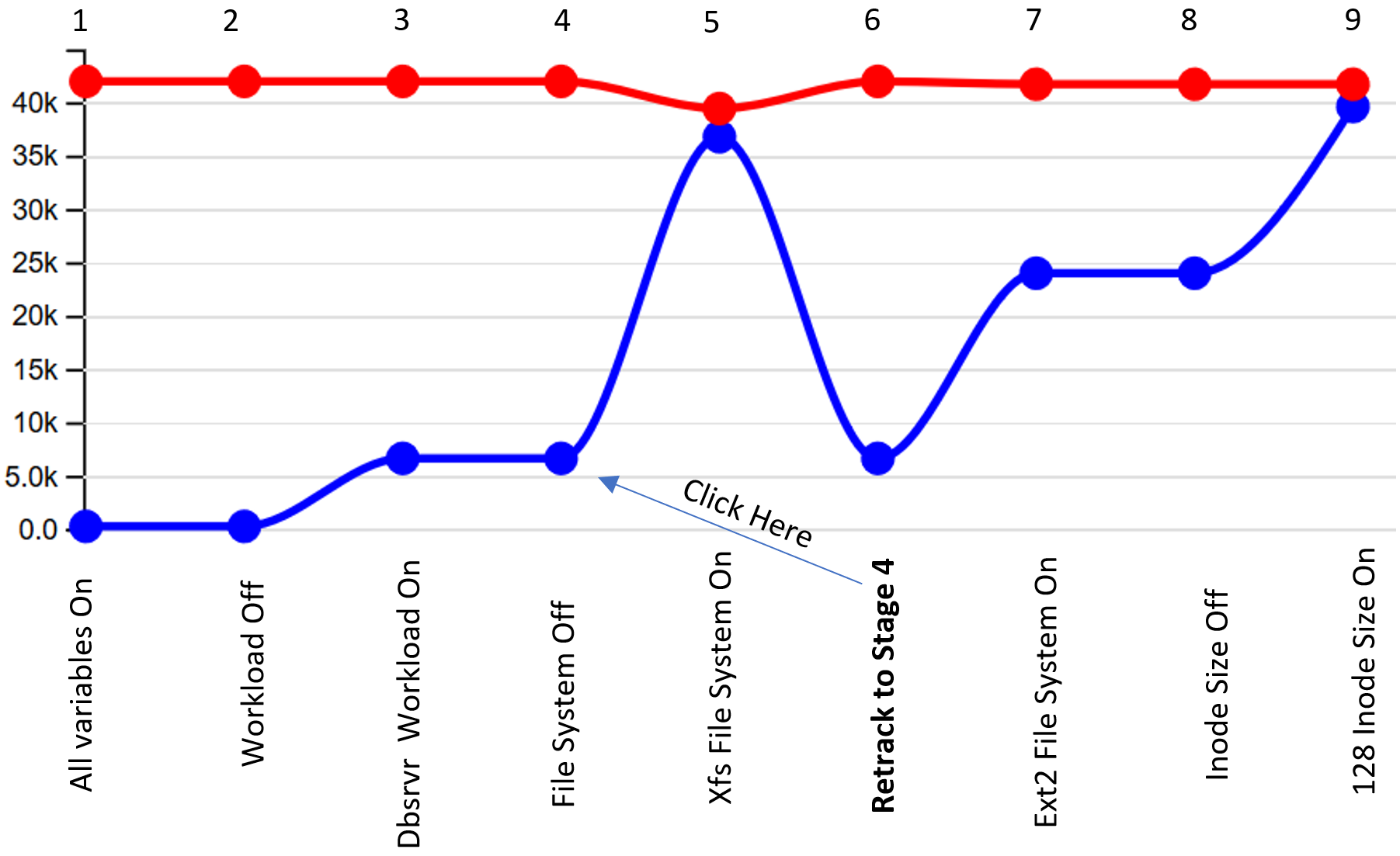}
 \caption{The Provenance Terminal in configuration filtering.  A system
   administrator is optimizing configurations for minimum throughput
   variation.  Stages 1--5 show parameter filtering for the configuration
   (Workload:Dbsrvr $\rightarrow$ FileSystem:Xfs).  This configuration
   achieves a minimum range of the throughput but the maximum throughput is
   reduced.  To get a better throughput, we check if choosing a different
   configuration after step 4 might help.  The user can roll back to step 4
   by clicking on the node (yielding step 6) and choose a different
   configuration (Workload:Dbsrvr $\rightarrow$ FileSystem:Ext2
   $\rightarrow$ InodeSize:128) shown in steps 7--9 to get a minimal range
   of throughput without compromising the maximum value.}
 \label{fig:timeline_interaction}
\end{figure}

An example use case of the Provenance Terminal can be that of a system
administrator searching for the best configuration but with a minimum
variation of the throughput.  The latter will reduce the uncertainty in the
predicted performance when the found parameter settings are applied in
practice.  The analyst would start off by selecting (Workload:Dbsrvr
$\rightarrow$ FileSystem:Xfs) as shown in stages 1--5 in
Figure~\ref{fig:timeline_interaction}.  We see that the minimum and the
maximum throughput values almost converge to a very small range, but the
maximum throughput value is compromised.  To correct this, the analyst can go
back to stage 4 by clicking on the red or blue pointer.  This leads to a
replication of this stage at the end of the chain as stage 6.  Now the
analyst can take a different path to get a better overall throughput while
simultaneously optimizing for minimum throughput range: i.e., stages 7--8 in
Figure~\ref{fig:timeline_interaction} (Workload:Dbsrvr $\rightarrow$
FileSystem:Ext2 $\rightarrow$ InodeSize:128).  In this way, the Provenance Terminal helps in comparing multiple configurations: i.e., comparing steps 1--5
(configuration 1) and steps 6--9 (configuration 2).

%% file: aggregate_view.tex
\subsection{Aggregate View}

The Aggregate View, located to the right of the Parameter Explorer
\textbf{B} in Figure~\ref{fig:teaser} displays a single R-D bar.
While the main purpose of each Parameter Explorer R-D bar is to convey the
dependent numerical variable distributions possible if the respective parameter level is
chosen, the Aggregate View communicates the distribution possible
with all currently selected parameter levels.  As such it can be used to
quickly visualize the impact of a transition from one parameter
configuration to another.  Whereas the Provenance Terminal summarizes the top and
bottom end of the achievable dependent variable's value only, the Aggregate View offers
detailed distribution information for the current parameter configuration.

%% file: interaction.tex
\subsection{Interaction with ICE: Two Case Studies}
To get a sense for how analysts would interact with ICE we present two use cases involving the systems performance dataset.  One practical
application is to analyze a system's performance stability.  Systems vary
greatly in their performance for different workloads which can be quantified by the aforementioned range, i.e., the difference between the maximum and the minimum
throughput for a particular configuration~\cite{cao2017performance}. A large range means less stability and less predictability.

\begin{figure}[tb]
 \centering
 \includegraphics[width=\columnwidth]{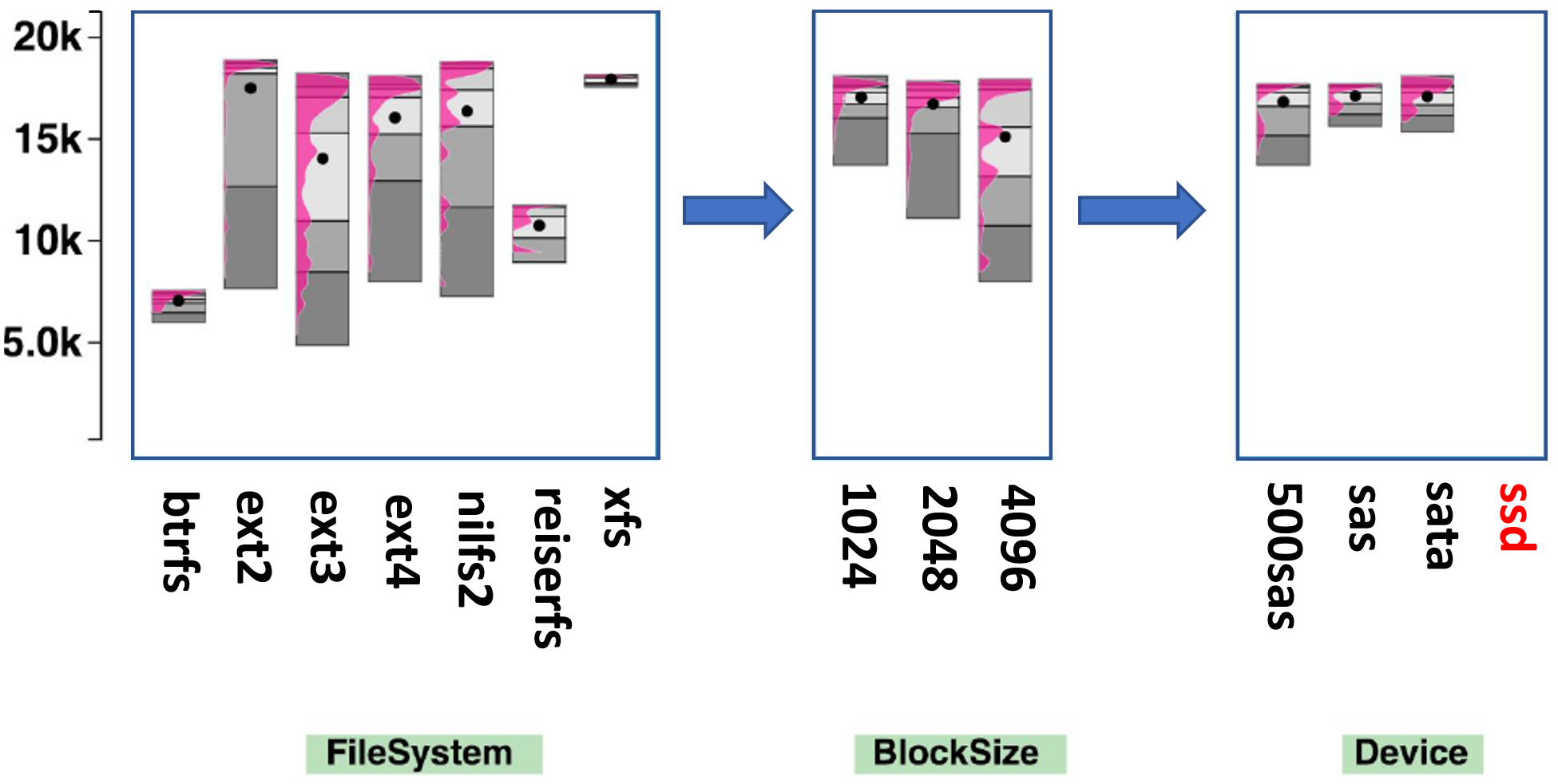}
 \caption{Using ICE to optimize a computer system running a mail server
   workload.  \textbf{Left to Right:} Three stages of the optimization
   process.}
 \label{fig:interaction1}
\end{figure}

The first use case shows how one would  optimize a system running a mail server workload. Figure~\ref{fig:interaction1} shows the steps involved in the filtering
process.  First, the analyst selects the workload type as Mail Server by clicking the respective label. The File System throughput values change as shown in the first
step in Figure~\ref{fig:interaction1}. The primary concern here is to minimize the variation in the throughput for a more stable and predictable mail service. The analyst can clearly see that choosing the \textit{btrfs} File System gives the minimum throughput range and thus is more stable and predictable for the user of the service. While its overall throughput is lower than for \textit{ext2} and \textit{ext4}, these File Systems are less reliable and would leave users of the mail service often frustrated.

However, sometimes there is a situation when the user cannot change the File System (i.e., because it requires a costly
disk reformat and restore), and thus it has to be set to \textit{ext4}
regardless of the application.  Such cases are quite common in practice, when it is not
possible to change some parameters of the system.  In such a case, the
analyst can return to the previous state of filtering by ways of the provenance terminal.
After selecting the \textit{ext4} File System, the next parameter to tune is
the block size which has throughput values as shown in Stage 2 of
Figure~\ref{fig:interaction1}.
Comparing the throughput distributions for each level in block size, the
user selects block size of 1024 since it results in the highest throughput
value with minimum variation.  After choosing Block Size = 1024, the parameter
explorer view is updated with new throughput distributions for each
parameter level.  The next parameter the user can filter is the device type,
shown as Stage 3 in Figure \ref{fig:interaction1}.  For the given
configuration, the device type \textit{ssd} cannot be chosen since there is
no sample with such configuration in the dataset.  The label is henceforth
colored red.  Now the analyst can select either a \textit{sas} or
\textit{sata} device.  This presents a trade-off where \textit{sas} has a
lower range while \textit{sata} gives a higher throughput.

%% file: design_alternatives.tex
\subsection{Design Alternatives}
\label{s:design-alt}

There were four design alternatives which we had to choose from.  In this
section we discuss why we chose the current design of the ICE tool given the
alternatives.

\begin{itemize}
\item \textbf{R-D bars instead of box plot:} Box plots are great for
  representing the distribution of data with the help of percentiles,
  but they show only fifty percent of the data (i.e., from 25$^{th}$ to
  75$^{th}$ percentile). They also assume that the data points are normally distributed which can be restrictive: it certainly is a restriction in our application as is apparent in the distributions shown in any of the R-D bars.

\item \textbf{R-D bars instead of parallel sets:} Bars make it possible to
  represent the parameters and their levels in a smaller space as compared
  to parallel sets.  The R-D bars also prevent data cluttering because they
  capture the configuration statistics succinctly without the need to draw
  individual lines (see also Section~\ref{param-exp}).
\item \textbf{Displaying the distribution:} Violin
  plots~\cite{hintze1998violin} and bean plots~\cite{kampstra2008beanplot}
  are better in displaying distributions, as opposed to box plots.  We
  choose to display only one half of the violin plots inside of the R-D bars
  because it better utilizes the bar real estate.  This is important since
  there might be a large number of parameters and so the width available to
  each bar is limited.  In the interest of accommodating more parameter
  levels in a uniform looking display, the system experts suggested that
  half-violin plots inside the bars were a better design.

  \item \textbf{Choice of colors:} The color choices for percentiles and the
    distribution on the R-D bars were decided with a user study.  In an
    interactive session, the system researchers were presented with several
    possible color combinations for the R-D bars chosen from color
    brewer~\cite{harrower2003colorbrewer}.  The present selection of colors
    were deemed most appropriate by the experts in terms of visual
    interpretation.
\end{itemize}

%% file: implementation.tex
\section{Implementation}
\begin{figure}[tb]
 \centering
 \includegraphics[width=\columnwidth]{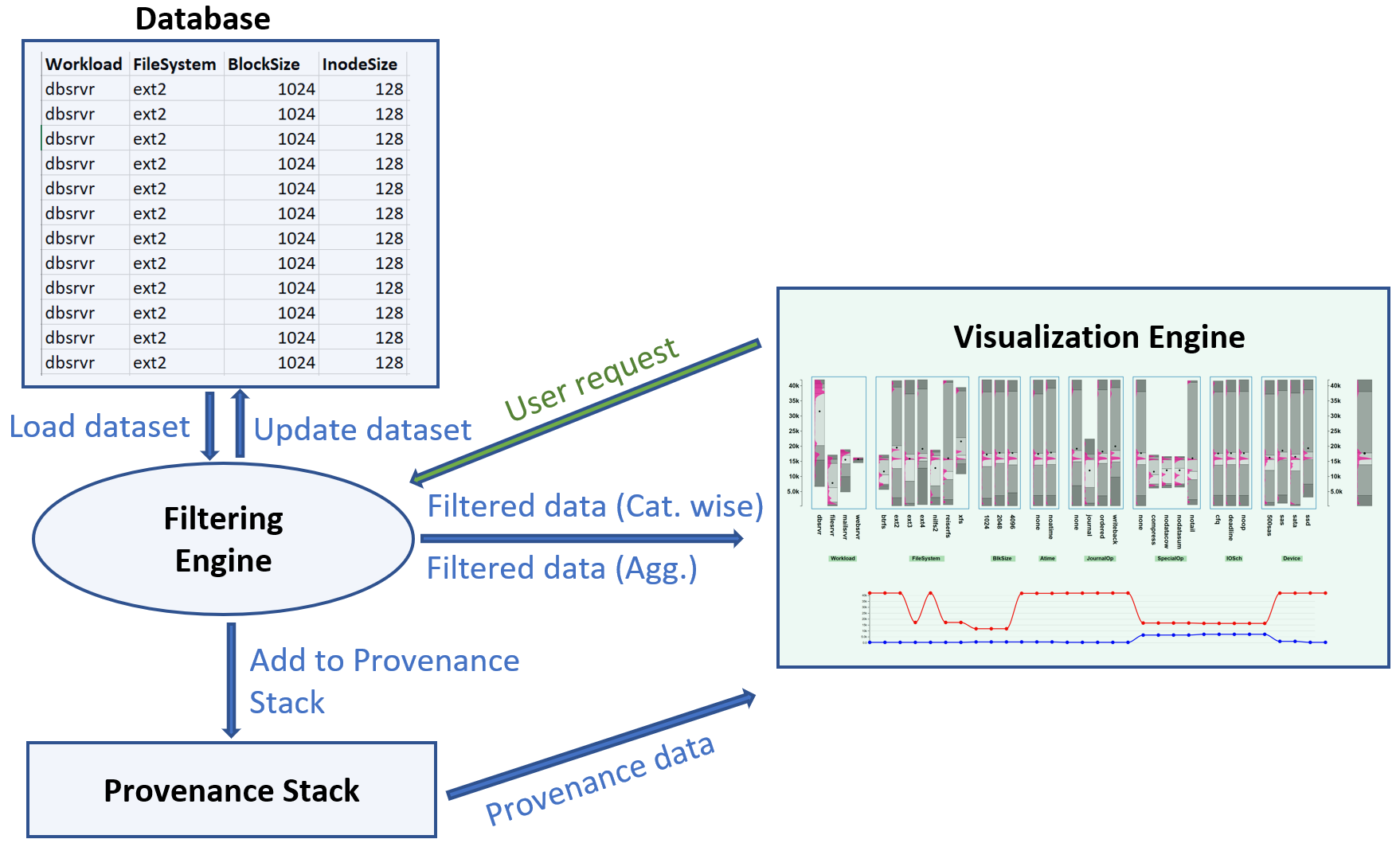}
 \caption{Block diagram showing the implementation of the ICE tool.  Green
   shows the requests handled by the Visualization engine and Blue shows the
   requests handled by the backend components i.e.  Filtering Engine,
   Dataset and the provenance stack.}
 \label{fig:block_diagram}
\end{figure}

Figure~\ref{fig:block_diagram} shows the block diagram of different
components of our ICE tool.  There is a backend server consisting of a
Database, Filtering Engine, and a Provenance Stack.  The frontend consists of
a Visualization Engine which runs in a browser.  The backend is a python flask server and the
frontend is created with $D^{3}$~\cite{bostock2011d3}.  A database
stores the original dataset which can be uploaded from the ICE interface.

The Filtering engine updates the existing data based on a user request from
the Visualization engine.  The data is then grouped separately for the
Parameter Explorer and the Aggregate View and sent to the Visualization
engine for display.  Another component to the backend is the Provenance Stack,
which keeps track of the dependent variable values with each user request.
With every interaction, the Filtering engine updates the Provenance Stack
which then updates the Provenance Terminal.

\subsection{Data filtering}

To filter and display large amount of data in real time is
challenging.  ICE is optimized for filtering speed using one-hot
encoding filtering and random sampling.  One-hot encoding is used to
convert categorical data to binary variables for faster processing with no
loss of information.  An example of converting the categorical data to numerical with one-hot encoding is provided in the supplementary material. This
technique greatly reduces the time complexity of searching for a parameter level. Where regular searching for a categorical parameter level
has $O(NM)$ complexity, one-hot encoding has $O(N)$ time complexity
($N$ is the number of datapoints and $M$ is the number of parameter levels).  Another benefit of using one-hot encoding is that it generates a
sparse version of the dataset which is easier for the modern systems to process with specialized data structures~\cite{golub2012matrix,
  tewarson1973desirable, pissanetzky1984sparse, duff2017direct}.

\begin{figure}[tb]
 \centering
 \includegraphics[width=\columnwidth]{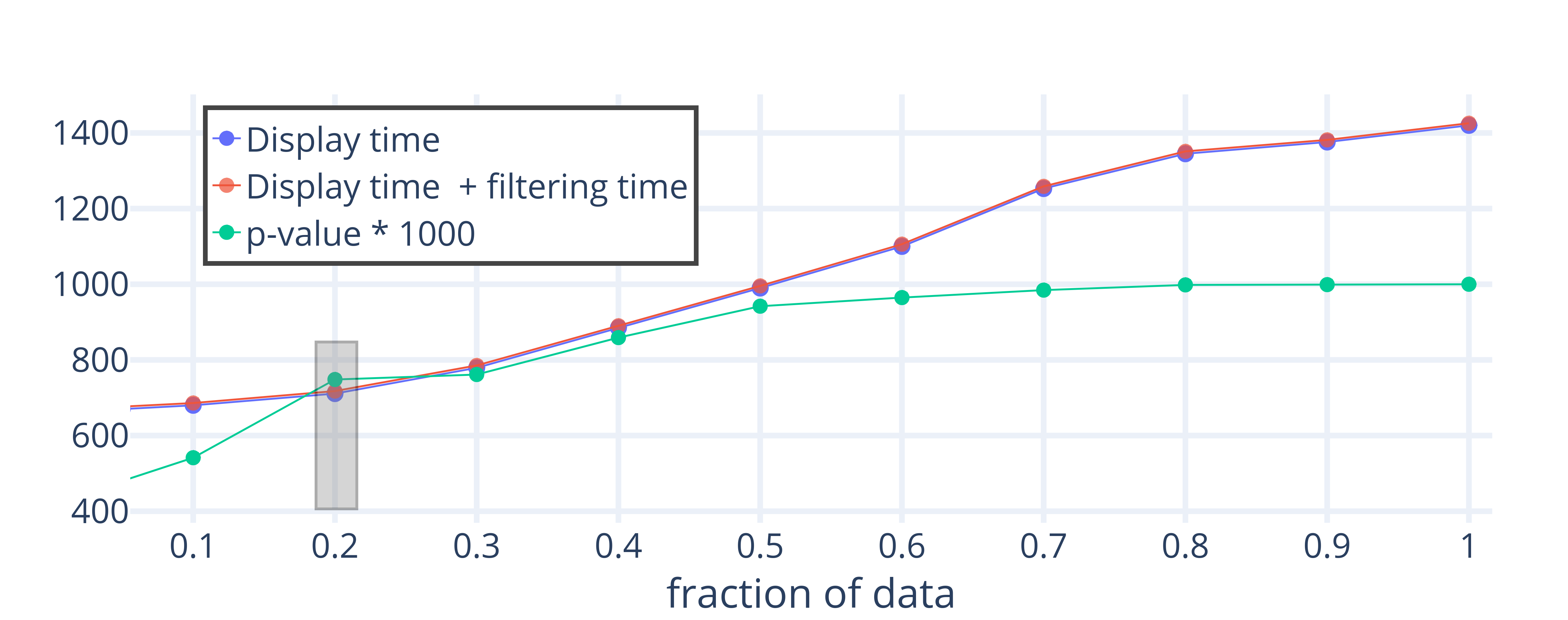}
 \caption{Observations of information loss with sampling and time to
   generate the visualization on the ICE tool.  Filtering time is much
   smaller, hence the orange and blue lines almost overlap. Note that the Y axis starts at 400.}
 \label{fig:time_pvalue}
\end{figure}

For the requirement to display distribution curves for each parameter level,
the time to display the filtered data also needs to be optimized.  If we try to
use every datapoint in the calculation of the distribution, the time to
display the visualization would not scale well with the size of the dataset.
The time to display full data on our dataset with around 100k configurations
is around 1,400 milliseconds, which is too slow.
Hence,
sampling of the data is required to estimate distributions.  We
evaluated the trade-off between information loss with random sampling and the time
to display the data.  Figure~\ref{fig:time_pvalue} shows that as the
distribution similarity (p-value) of the complete and sampled dataset
increase, the time to generate the visualization also increase. To measure information loss with sampling, we used the Kolmogorov-Smirnov test by comparing data distribution from the sampled dataset with the complete dataset.  


After evaluating the loss of information with sampling and the time to
display the visualizations, a sample size of 20\%
proved to be an appropriate option.  This is because the display time curve has a steep increase as we go to higher sample sizes but the the p-value does not
increase much after 20\%---hence a good trade-off. ICE on the systems performance dataset uses 20\% of the full dataset (20k data points) which takes around 800 milliseconds
of display and filtering time. These results also give a good threshold for dataset size which can be fully displayed with ICE without sampling. In the current implementation of ICE, the datasets with less than 20k data points are processed without sampling. For larger datasets, the sample size is determined when the p-value crosses a .5 threshold.

%% file: evaluation.tex
\section{Evaluation}

In this section, we evaluate our ICE using the techniques suggested in the
nested model-based visualization design literature~\cite{munzner2009nested,
  meyer2012four}.  We first used the Analysis of Competing Hypotheses
(ACH)~\cite{heuer1999analysis} method as a mechanism to efficiently identify
which of the existing techniques (see Section~\ref{s:related}) would need to
be formally compared with ours via a user study.  The ACH is a methodology
for an unbiased comparison of a set of competing hypotheses, in our case the
various visualization techniques in terms of the requirements put forward in
Section~\ref{s:req}.

The ACH showed that only ICE and Parallel Sets could satisfy all formulated
hypotheses.  We did not consider hypotheses comparing the goodness of a
visualization or the effectiveness of filtering as these could be improved
in any existing technique.  Also, determining the goodness of a
visualization is difficult~\cite{johnson2005nih} and requires a subjective
study.  We then conducted a formal user study to compare Parallel Sets with
ICE.


\subsection{Initial Comparative Evaluation Using ACH}

The Analysis of Competing Hypotheses (ACH) is a technique to choose the best
possible solution to satisfy a set of hypotheses.  Fitting our overarching
application scenario, we only evaluated the existing techniques (and ICE) in
terms of the specific task of analyzing a set of categorical data with
respect to a numerical target variable.  It corresponds to the interaction
and technique design stage of the nested model by Munzner
\textit{et. al}~\cite{munzner2009nested, meyer2012four}.  We derived six
hypotheses from the requirements listed by the system performance experts
(see Section~\ref{s:req}) as follows:

\begin{table}[tb]

  \caption{Competing hypotheses analysis on existing visualization
    techniques and our ICE tool.  A check mark means the hypotheses is
    satisfied whereas a cross mark means that the hypotheses is not satisfied
    by the given visualization technique.  The results at the bottom shows
    the accepted visualization techniques (i.e., satisfy all the hypotheses)
    and the rejected ones (do not satisfy at least one of the hypotheses).
  \label{tab:comp_hyp}}
  \scriptsize%
	\centering%
  \begin{tabu}{%
	r%
	*{7}{c}%
	*{2}{r}%
	}
  \toprule
  \rotatebox{90}{Hypotheses} & \rotatebox{90}{Fused Displays} \rotatebox{90}{MCA} & \rotatebox{90}{\textbf{Parallel Sets}} & \rotatebox{90}{Dimension reduction} \rotatebox{90}{MDS, T-Sne,} \rotatebox{90}{LDA, LLE} \rotatebox{90}{Isomap,} \rotatebox{90}{Spectral Clustering} & \rotatebox{90}{Cramer's V and} \rotatebox{90}{Scatterplot Matrix} & \rotatebox{90}{Bi-plots} & \rotatebox{90}{\textbf{ICE tool}}\\
  \midrule
  H1 & \checkmark & \checkmark & $\Cross$ & \checkmark & $\Cross$ & \checkmark \\
  \midrule
  H2 & \checkmark & \checkmark & $\Cross$ & \checkmark & \checkmark & \checkmark \\
  \midrule
  H3 & $\Cross$ & \checkmark & $\Cross$ & $\Cross$ & $\Cross$ & \checkmark\\
  \midrule
  H4 & $\Cross$ & \checkmark & \checkmark & $\Cross$ & \checkmark & \checkmark \\
  \midrule
  H5 & \checkmark & \checkmark & $\Cross$ & \checkmark & $\Cross$ & \checkmark \\
  \midrule
  H6 & $\Cross$ & \checkmark & \checkmark & $\Cross$ & \checkmark & \checkmark \\
  \midrule
  \textbf{Result} & $\Cross$ & \checkmark & $\Cross$ & $\Cross$ & $\Cross$ & \checkmark \\
 \bottomrule
  \end{tabu}%
\end{table}

\textbf{H1: Allow an assessment of the distribution of a numerical variable
  in terms of a given parameter.} The visualization is able to display the
distributions of the dependent numerical variable for each parameter.  The
analyst can get an estimate of the nature of this distribution:
bi-modal, multi-modal, uniform, normal distributed, etc.

\textbf{H2: Allow an assessment of the correlation between parameters.} The
visualization makes it possible to compare or correlate the parameters in
the dataset with respect to their impact on the target numerical variable.
Irrespective of the method of correlation, the analyst should be able to
derive informative conclusions while filtering the parameter space based on
correlation.

\textbf{H3: Enable quick filtering.} Filtering is used to track the best
performing configurations for a desired goal.  The visualization technique
enables the analyst to add, remove and edit the parameters of the
configuration and see updated distribution of the dependent numerical
variable within one second.

\textbf{H4: Allow an assessment of the statistics alongside the
  distribution.}  The visualization technique displays the statistics (mean,
median, percentiles, max, and min) of the dependent numerical variable for
each parameter.

\textbf{H5: Allow informed predictions.} The visualization provides cues to
the analyst for filtering the parameter space.

\textbf{H6: Provide insight on aggregate distributions.} Similar to
requirement R6, the visualization technique provides a summarized display of
the dependent numerical variable values which can be reached from a given
parameter setting.

We left out a hypothesis for the provenance visualization because it was not
supported by any of the existing techniques (only ICE).
Table~\ref{tab:comp_hyp} shows the results of the ACH-based evaluation
applied to the available visualization techniques and our ICE.
The comparison
shows that by eliminating any visualization technique which does not satisfy
one or more of the hypotheses, only parallel sets and ICE fit all
hypotheses.

\subsection{User Study Comparing Parallel Sets and ICE}

Although the ACH evaluation revealed that both Parallel sets and ICE could be
used to analyze categorical variables in the context of a target numerical
variable, our computer systems experts voted against the use of Parallel
Sets.  This was because Parallel sets become too cluttered to effectively
filter the parameter space for larger datasets.  Nevertheless, to make these
informal impressions more concrete, we conducted a user study to compare the
effectiveness of ICE and Parallel Sets.  The main objective of the user
study was to compare the ICE and Parallel Sets based on two metrics:
\textit{Time to filter configurations} and \textit{Accuracy of filtering}.
The participants in the user study were divided into three categories based
on their expertise: \textit{System performance experts (SE)},
\textit{Visualization experts (VE)}, and \textit{Non experts (NE)}.
\textit{SEs} were researchers working in the area of system performance,
\textit{VEs} were researchers working in the area of visual analytics, and
\textit{NEs} were users with no research experience in either of the two
areas.

A question bank for the user study was compiled with the questions designed
by three system researchers (independently), to uniformly represent the
requirements of the systems community.  After an initial usage tutorial,
participants were given two unique sets of five randomly sampled tasks
from the question bank to perform on both the tools.  The dataset used in the study was the systems performance dataset as described in Section \ref{s:dataset}. The user study was
conducted on 21 users: 7 \textit{SE's}, 7 \textit{VE's}, and 7 \textit{NE's}.
Among the total participants, the gender composition was 9 females and 12
males with the overall age range of the participants being 22 to 34 years.

The results of the user study proved the effectiveness of ICE tool over
Parallel Sets both in terms of accuracy and time to filter the parameter
space.  The average time for users to solve a question on ICE tool was 47.6
seconds as compared to Parallel sets which was 73.3 seconds.  To compare the
statistical significance of time difference, we performed a paired t-test on
the distributions of average time to answer a question for each user on both
the tools.  The p-value of the single tailed t-test was p = .0074 which is
lower than the significant value of .05.  Hence, the mean time to filter
the parameter space is lower in ICE as compared to Parallel Sets with a high probability.

A similar analysis was done to measure the accuracy of each user on the five
questions in the user study.  The average accuracy of the participants using
the ICE tool was 4.37 compared to 2.75 for parallel sets.  The p-value
obtained on the single tailed t-test for the comparing accuracy
distributions was p $<$ .001, which is significantly lower than the threshold of
.05.  Hence, the mean accuracy of the analyst for parameter filtering via
the ICE tool is higher than via the Parallel Sets with a high probability.  Given the results of this user study we conclude that ICE is
better for multidimensional parameter space analysis both in terms of
accuracy and time when compared to Parallel Sets.

We also analyzed the mean accuracy and time based on user expertise.  The
\textit{NEs} took the most time for answering the user study questions and
had the lowest accuracy as compared to other expertise categories with both
of the tools.  Also, the \textit{VEs} were the most accurate with their
answers but took a little more time compared to the \textit{SEs}.  However,
the trend of expertise-wise accuracy and time is the same for both ICE and
Parallel sets.  All plots for the expertise wise analysis and the user study
tasks along with the dataset are provided in the supplementary material.

\subsection{Case Studies}

We also evaluated the ICE with case studies derived from two datasets taken
from Kaggle.com~\cite{HRdata, FrenchPopulation}.  One dataset is an HR
dataset of a US firm containing data on the hourly pay of its employees
based on various parameters.  The other is a French population
characteristics dataset where the population distribution of a set of cities
in France is studied on the basis of gender, cohabitation type, and age
groups.  Two domain experts were consulted to evaluate the effectiveness of
our ICE tool in the study of different parameters in these datasets.  Expert
A who evaluated the ICE tool on the HR dataset had management experience at
a private firm, and Expert B who evaluated the ICE tool on the French
population dataset was an expert survey analyst.

\subsubsection{Exploring the HR Dataset}
\label{s:eval:exp}

\begin{figure}[tb]
 \centering
 \includegraphics[width=\columnwidth]{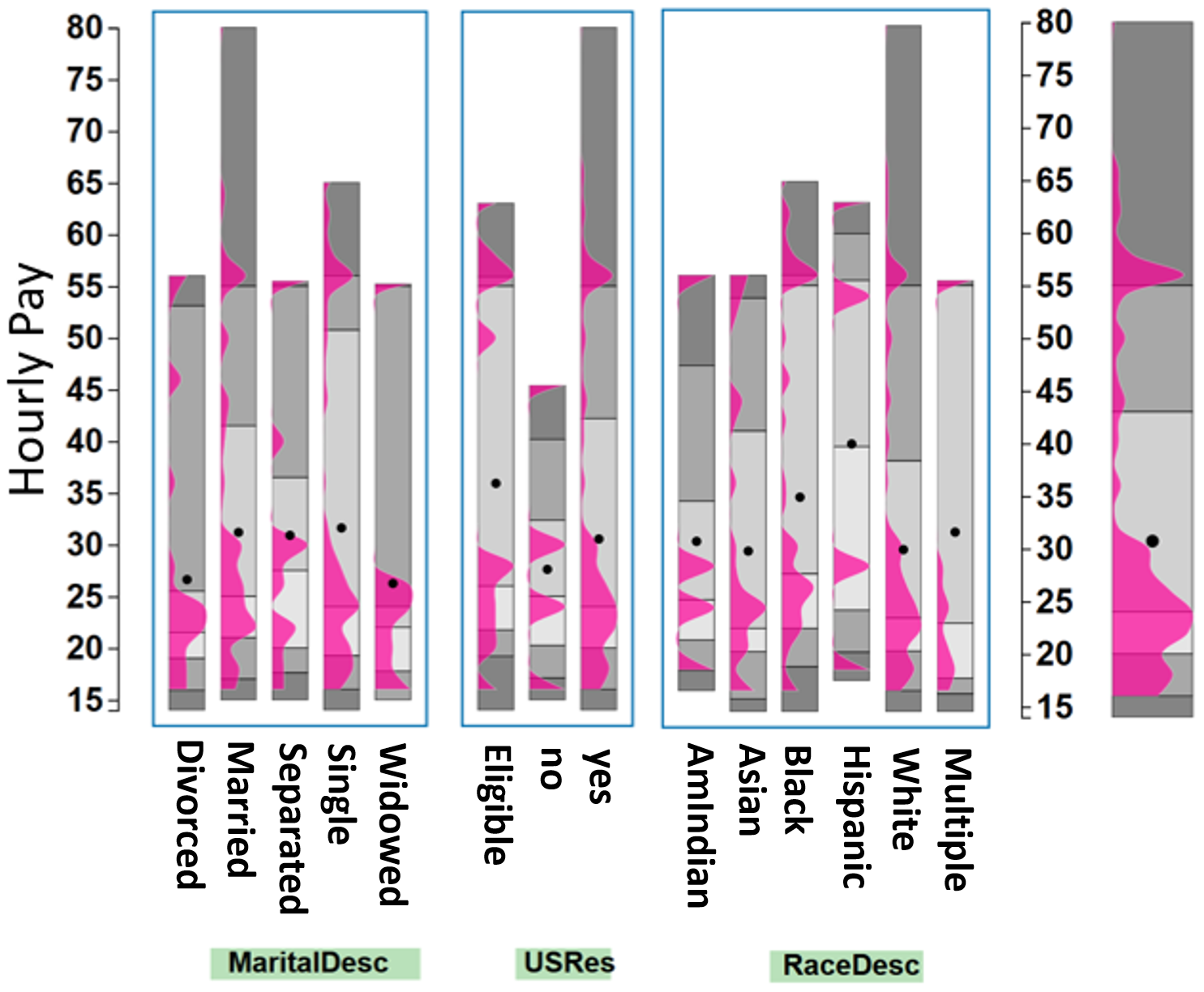}
 \caption{Using the ICE to explore the HR dataset from Kaggle.com.  There
   are seven variables in the dataset (Figure shortened due to lack of
   space); the numerical dependent variable is hourly pay scale.}
 \label{fig:hr_dataset_1}
\end{figure}

This case study uses the ICE for exploring the HR dataset.
The dataset has seven categorical variables: (\textit{Marital Status, US
  Residency Status, Hispanic status, Race, Department, Employee Status and
  Performance Score}) and one dependent numerical variable: \textit{Hourly
  Pay Rate}.  To start out, Expert A (EA) first familiarized himself with
the dataset and the usage of the ICE tool.  Figure~\ref{fig:hr_dataset_1}
has a part of the initial screen he browsed.  It shows three of the seven
variables with respect to hourly pay scale.  Some of the more interesting
observations he made
were: (1) Married workers had the highest hourly pay and the mean hourly pay
was highest for single workers.  (2) The mean hourly pay of non-residents
who are eligible for US citizenship is higher than those of the residents.
(3) White workers have the highest hourly pay among all races.  (4)
Considering the departments, the executive department had the highest hourly
pay scale followed by IT services.

\begin{figure}[tb]
 \centering
 \includegraphics[width=\columnwidth]{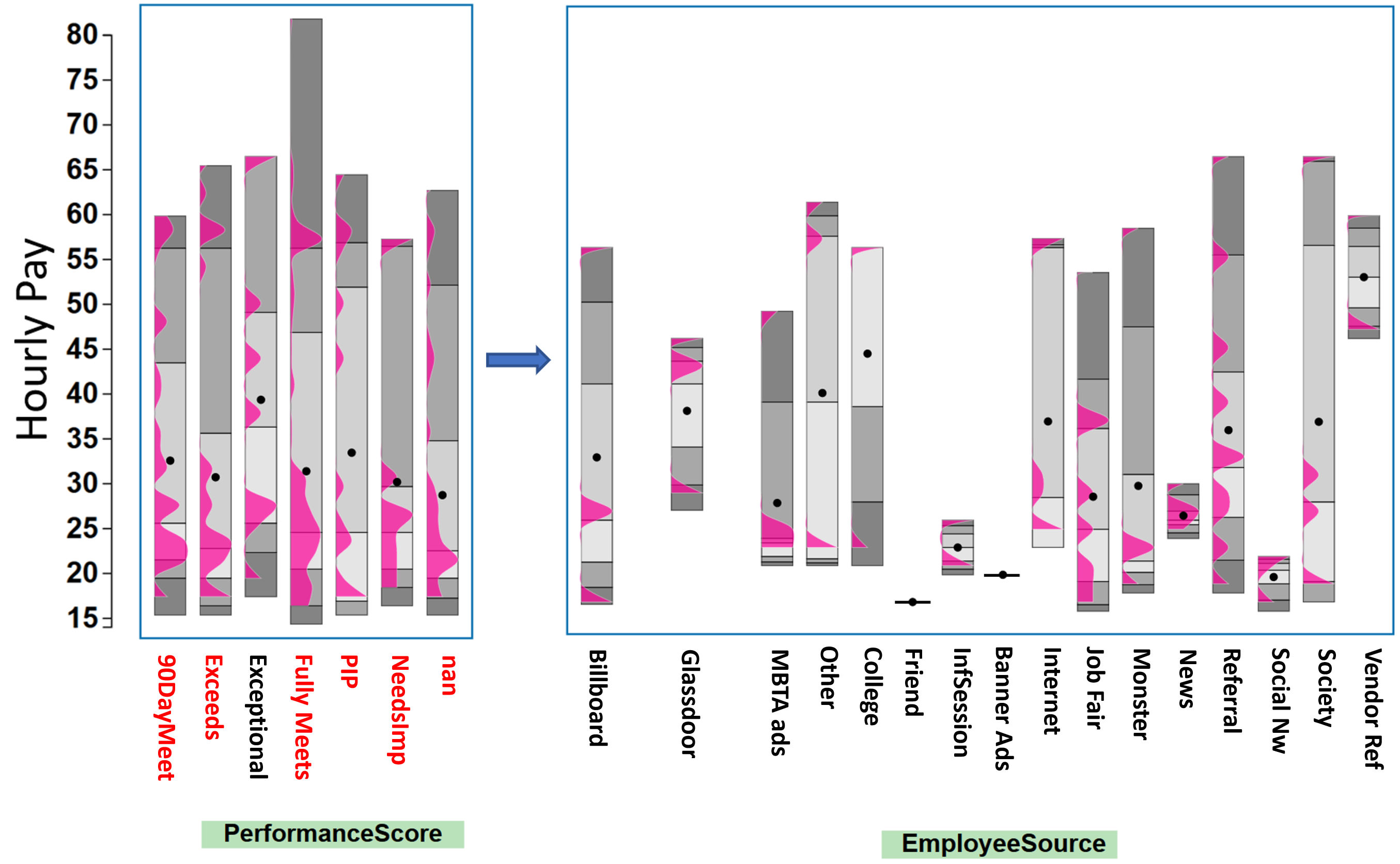}
 \caption{Using the ICE to explore the employee source of workers with
   exceptional performance.  (Left panel) Selecting the performance score as
   exceptional.  (Right panel): The filtered employee source parameter group
   where many interesting observations can be made (see text in
   Section~\ref{s:eval:exp}).  For example, it appears that the exceptional
   employees hired from vendor referral have the highest mean hourly pay and
   with a fairly low range.  This probably is because these individuals had
   to be paid at competitive rates to make the switch.  Conversely, passive
   advertising such as billboard ads, monster, and news also yielded many
   exceptional employees but at lower pay overhead.  Finally, college and
   information sessions yielded the lowest number of exceptional employees.}
 \label{fig:hr_dataset_2}
\end{figure}

After the initial analysis, the other two variables in the dataset that were
of particular interest to EA were Employee Source and Performance Score.  He
wanted to see whether high performing employees were properly compensated
for their valuable efforts.  The Parameter Explorer made this investigation
easy and EA quickly confirmed that exceptional employees were indeed paid
more than other employees, with a mean pay of about \$40 per hour, shown in
Figure~\ref{fig:hr_dataset_2}.

Another parameter of interest was the hiring source of these exceptional
employees.  EA selected the exceptional performance score in the Parameter
Explorer.  This filtering updated the Employment Source group to only show
the sources of exceptional workers with respect to their hourly pay.
Figure~\ref{fig:hr_dataset_2} shows the result of this filtering and the
caption offers a few interesting observations.


EA suggested that for better equality of all sources of exceptional workers,
their mean hourly pay should be similar.  Also, EA suggested that investment
on college fairs and job sessions should be lowered as they are not a good
source of exceptional workers.  EA then confirmed that the use of ICE would
help the HR department to better manage the company's funding and
investments.

\subsubsection{Exploring the French Population Dataset}

\begin{figure}[tb]
 \centering
 \includegraphics[width=\columnwidth]{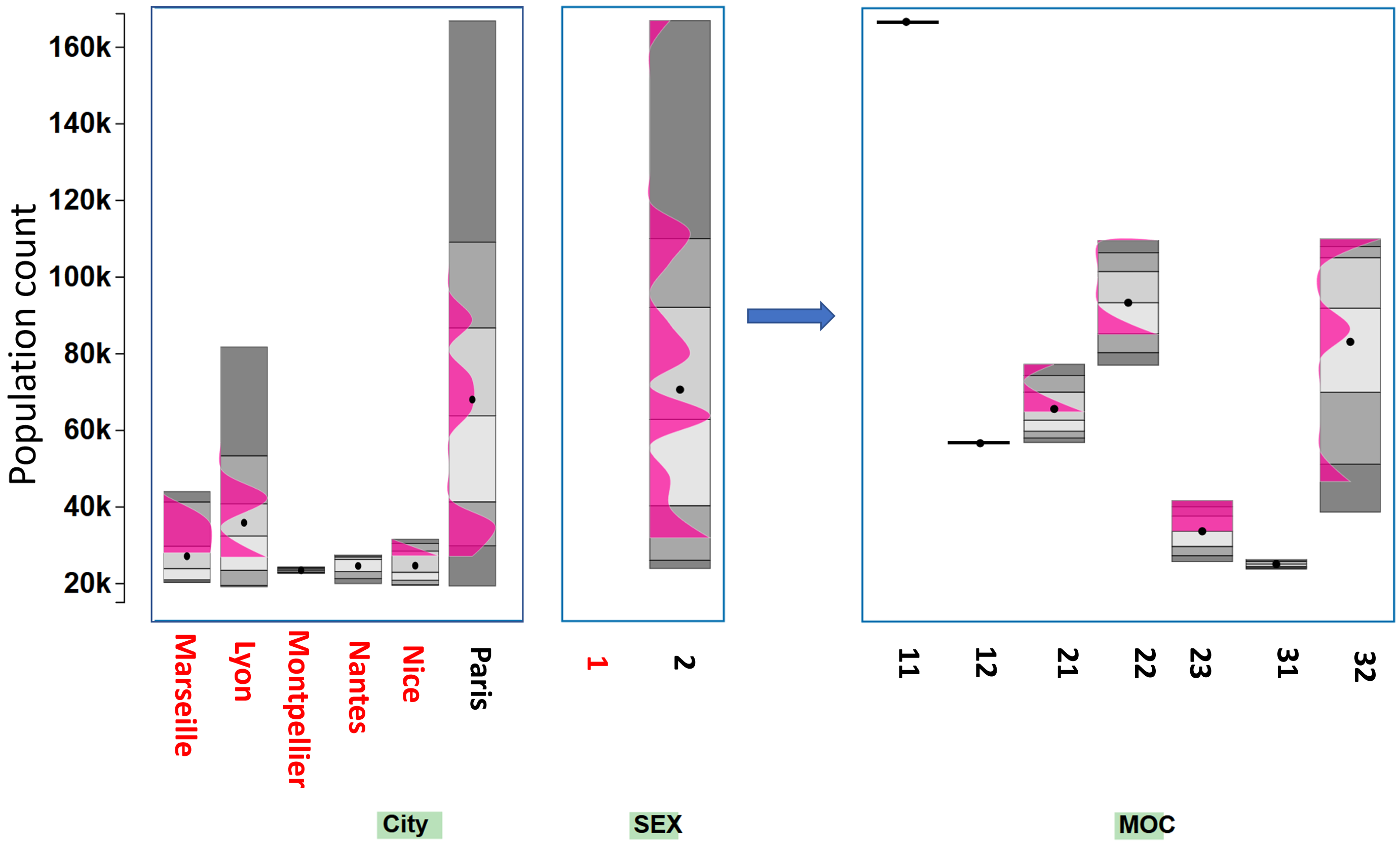}
 \caption{The ICE tool visualizing the French population habitation
   characteristics for city=Paris and gender type=2 (Female).  The figure
   shows three variables: City Name, MOC (Method of cohabitation) and the
   sex of the population distribution.  MOC
   numbers mean, 11: children living with two parents, 12: children living
   with one parent, 21: Adults living in couple without child, 22: Adults
   living alone with children, 31: person not from family, 32: persons
   living alone.  Age is described in groups where a number 20 means the age
   from 0 to 20, 40 means age group of 21 to 40 and likewise.}
 \label{fig:france_dataset}
\end{figure}

The French population habitation dataset has been collected to show existing
equalities and inequalities in France.  It consists of four categorical
variables (\textit{City, MOC (Method of Cohabitation), Age group, and sex}).
The dependent numerical variable, \textit{Population count}, is the number of people in
each of the categories defined by permutations of the independent variables,
for example, one category might be adult females with age 21--40 living in
Paris with her children.  Expert B (EB) was a survey analyst and like Expert
A he first familiarized himself with the ICE tool by looking at an overview
of the dataset's variables.  The overview screen of the ICE showing the
population distributions and statistics is provided in the supplementary
material.  EB's initial observations were: (1) The population count for a
few categories in Paris is exceptionally high compared to other categories
because the mean is very low compared to the highest value.  This can also
be seen in Figure~\ref{fig:france_dataset}. (2) The mean population of the
age range 60--80 is the highest in all cities; (3) The age group 20 to 40 is
the lowest on average for all cities; and (4) The average number of females
is higher than the average number of males for the overall population.

Following the basic inferences, EB was further interested to study the
habitation methods of females in three major cities of France: Paris,
Marseille, and Lyon.  EB selected Paris from the City variable followed by 2
from the Gender variable.  The Parameter Explorer then showed the
distributions of population for all categories of habitation methods, as
shown in Figure~\ref{fig:france_dataset}.  EB could see that the most
females were children living with two parents, i.e., category 11 (shown by a
single dot because all of these females have the same age group of below 20
years) followed by females living alone (i.e., category 32).  Similar
analyses were done for the cities of Marseille and Lyon.  For Marseille, EB
pointed out that almost an equal number of females lived as a single
household and in a family with children.  For Lyon, most females were the
children living with two parents followed by females living as a single
household.  EB then used the Provenance Terminal to go back two stages in the
filtering process to compare the female habitations in all cities.  EB
further pointed out that Paris had exceptionally large number of children
living with single parent as compared to other cities.

After evaluating the use of the ICE on the France population dataset, EB
recommended ICE as an effective tool for the quick filtering and
understanding of survey statistics.  EB also found the ICE tool helpful in
understanding biases in the population distributions.

%% file: conclusion.tex
\section{Conclusions}

This paper presents the ICE tool, a novel approach for categorical parameter
space analysis in the context of a dependent numerical variable.  ICE
overcomes the existing challenges by providing an effective layout for
parameter space visualization.  The stacked R-D bars concept used in ICE
along with interaction assists in effective filtering of the parameter
space.  A greater number of parameters could be visualized and readily
correlated, thus increasing the efficiency of filtering.  Multiple
configurations can be compared for their impact on the target variable based
on any objective.  ICE also supports multi-objective filtering since it
presents full statistics and distribution information to the user for each
parameter level.

Several important lessons were learned while designing the idea of ICE.  In
the requirement analysis phase with the systems community researchers, we
realized that by presenting the results gathered from the dataset with
existing visualization techniques helped make the gathering of requirements
more effective.  Almost from the start, the system experts were skeptical
about the accuracy of most existing techniques.  They wanted a tool that
would be able to show the statistical distributions precisely.  It also
helps to keep a keen eye on any struggles the collaborating domain experts
may experience.  For example, in the filtering experiments we noticed that
they had trouble remembering the filtering path.  This gave rise to the
provenance terminal.

Besides the effective design of ICE, there still remain some limitations
which can be taken up as the future work.  For larger datasets, techniques
to combine multiple parameters~\cite{keim2008visual} can be incorporated to
prevent excessive thinning of the bars. Moreover, some related precomputed solutions can be provided to the analyst based on optimization objectives to start off with the search process. Also, ICE is based on the
assumption that the cost of changing parameters is the same throughout,
which might not be true in some cases.  Moreover, these costs might vary
with time~\cite{van2017automatic}.  It will be useful to incorporate cost
measures into ICE and provide support for real time cost based filtering.
We will continue working on our ICE tool to incorporate these new features.



%% file: ack.tex
\section*{Acknowledgments}
\label{sec:ack}

We would like to thank the anonymous VAST 2019 reviewers for their
valuable comments.
This work was made possible in part thanks to Dell-EMC, NetApp, and
IBM support; NSF awards IIS-1527200, CNS-1251137, CNS-1302246, CNS-1305360,
CNS-1622832, CNS-1650499, and CNS-1730726; and ONR award
N00014-16-1-2264.
